\newcommand{\cali}{\mathcal{I}}

\newtheorem{definition}{Definition}

\newtheorem{lemma}{Lemma}
\newtheorem{property}{Property}

\documentclass[journal]{IEEEtran}

\usepackage{graphicx}

\begin{document}
%
\title{A Dynamic Programming Approach To Length-Limited Huffman Coding}
%
%
%

\author{Mordecai~Golin,~\IEEEmembership{Member,~IEEE,}
        and~Yan~Zhang
\thanks{M. Golin and Y. Zhang are with the
Department of Computer Science \& Engineering,  Hong Kong UST,
ClearWater Bay, Kowloon, Hong Kong. Their research was partially supported by HK RGC CERG grants
HKUST6312/04E and 613105.}
}


%

\maketitle

\begin{abstract}
The ``state-of-the-art'' in Length Limited Huffman Coding algorithms is the 
$\Theta(ND)$-time, $\Theta(N)$-space one of Hirschberg and Larmore, where $D\le N$ is the length restriction
on the code.  This is a very clever, very problem specific,
technique.  In this note we show that there is a simple Dynamic-Programming (DP) method
that solves the problem with the same time and space bounds.  
The fact that there was an
$\Theta(ND)$ time DP algorithm was previously known; it is a 
straightforward DP with the {\em Monge} property  (which permits an order
of magnitude speedup).  It was not
 interesting, though, because it also required
$\Theta(ND)$ space.

The main result of this paper is 
the {\em  technique} developed for reducing the space. It is quite simple and applicable to many
other problems modeled by DPs with the Monge property.  
We illustrate this with examples from web-proxy design and wireless mobile paging.
\end{abstract}

\begin{IEEEkeywords}
Prefix-Free Codes,  Huffman Coding,  Dynamic Programming,  Web-Proxies,  Wireless Paging, the Monge property.
\end{IEEEkeywords}

%
\IEEEpeerreviewmaketitle

\section{Introduction}
\label{sec:intro}
\IEEEPARstart{O}{}ptimal prefix-free coding, or \emph{Huffman coding}, is a standard compression technique.
Given an \emph{encoding alphabet}
$\Sigma = \{\sigma_1, \ldots, \sigma_r\}$, a {\em code} is just a set of words in $\Sigma^*.$
Given $n$ probabilities or nonnegative frequencies $\{p_i : 1 \le i \le n\}$, and associated code $\{w_1,w_2,\ldots, w_n\}$
the {\em cost} of the code is $\sum_{i=1}^n p_i |w_i|$ where $|w_i|$ denotes the length of
$w_i$.
A code is {\em prefix-free} if no codeword $w_i$ is a prefix of any other codeword $w_j$.
An {\em optimal} prefix-free code for $\{p_i : 1 \le i \le n\}$
is a prefix-free code that minimizes its cost among all prefix-free codes.

In \cite{Hu52}, Huffman gave the now classical $O(n \log n)$ time algorithm for solving this problem.
If the $p_i$'s are given in sorted order, Huffman's algorithm can be improved to $O(n)$ time
\cite{Va76}.
In this note we will always assume that the $p_i$'s are presorted and  that $p_1 \le p_2 \le \ldots \le p_n$.

In some applications, it is desirable that the length of all code words are bounded by a constant,
i.e., $|w_i| \le D$ where $D$ is given. The problem of finding the minimal cost prefix-free code among
all codes satisfying this length constraint is
the \emph{length-limited Huffman coding} (LLHC) problem, which we will consider here.
Fig.~\ref{fig:seq_ex} gives an example of inputs for which the Huffman code is {\em not} the same as the length-limited Huffman code.

The first algorithm for LLHC was due to Karp \cite{Ka61} in 1961; his
algorithm is based on integer linear programming (ILP), which, using standard ILP solving techniques,
leads to an exponential time algorithm.
Gilbert \cite{Gi71}
in 1971 was interested in this problem because of the issue of inaccurately known sources;
since the probabilities $p_i$'s are not known precisely, a set of codes with limited length will,
in some sense, be ``safe''.  The algorithm presented in \cite{Gi71} was an enumeration one
and therefore also runs in exponential time.
In 1972 Hu and Tan \cite{HuTa72} developed an $O(nD 2^D)$
time Dynamic Programming (DP) algorithm.
The first polynomial time algorithm, running in $O(n^2 D)$ time and using
$O(n^2 D)$ space, was presented by Garey in 1974 \cite{Ga74}.
Garey's algorithm was based on a DP formulation similar to that developed by
Knuth for deriving optimal binary search trees in \cite{Kn71} and hence only works for
binary encoding alphabets.
A decade later, Larmore \cite{La87} gave an algorithm running in $O(n^{3/2} D \log^{1/2} n)$ time
and using $O(n^{3/2} D \log^{-1/2} n)$ space.
This algorithm is a hybrid of \cite{HuTa72} and \cite{Ga74}, and therefore also only works for the
binary case.
This was finally improved by
Larmore and Hirschberg \cite{LaHi90} who gave a totally different algorithm
running in $O(nD)$ time and using $O(n)$ space.
In that paper, the authors first transform the length-limited Huffman coding problem to the
\emph{Coin Collector's} problem, a special type of Knapsack problem, and then, solve
the Coin Collector's problem by  what they name the \emph{Package-Merge} algorithm.
Their result is a very clever special case algorithm developed for this specific problem.

Theoretically, Larmore and Hirschberg's result was later superseded for the case\footnote{
$f(n) = \omega(g(n))$ if $\exists N,c > 0$ such that $\forall n >N$, $f(n) \ge g(n).$}
$D = \omega(\log n)$
by two algorithms based on
the \emph{parametric search} paradigm \cite{Me83}.
The algorithm by Aggarwal, Schieber and Tokuyama \cite{AgSc94+} runs in
$O(n \sqrt{D \log n} + n \log n)$ time and $O(n)$ space.  A later improvement by Schieber \cite{Sc98} runs
in $n 2^{O\left(\sqrt{\log D \log\log n}\right)}$ time and uses $O(n)$ space.
These algorithms are very complicated, though, and even for
$D = \omega(\log n)$, the Larmore-Hirschberg one is the one used in practice
\cite{TuMo96,TuMo95}.
For completeness, we point out that the algorithms of \cite{LaHi90,AgSc94+,Sc98} are all only
claimed for the binary ($r = 2$) case but they can be extended to work for the
non-binary ($r > 2)$ case using observations similar to those we provide in Appendix \ref{app:LLHCDP}
for the derivation of a DP for the generic $r$-ary LLHC problem.

Shortly after \cite{LaHi90} appeared, Larmore and Przytycka
\cite{LaPr91,larmore:1163}, in the context of parallel programming,
gave a simple dynamic programming
formulation for the binary Huffman coding problem.  Although their DP
was for regular Huffman coding and not the LLHC problem, we will see that it is quite
easy to modify their DP to model the LLHC problem.
It is then straightforward to show that their formulation also permits constructing
the optimal tree in $\Theta(nD)$ time by constructing a size $\Theta(nD)$ DP table.  
This is done is Section \ref{Sec:DP}.
This straight DP
approach would not be as good as the Larmore-Hirschberg one, though, because, like
many DP algorithms, it requires maintaining the entire DP table to
permit backtracking to construct the solution, which would require $\Theta(nD)$ space. The main
result of this  note is the development of 
 a simple technique (section \ref{Sec:space})
that permits reducing the DP space consumption down to $O(n)$,
thus matching the Larmore-Hirschberg performance with a straightforward DP model.
Our technique is not restricted to Length-Limited coding.  It can be used to reduce space
from $O(nD)$ to $O(n+D)$ in a variety of $O(nD)$ time DPs in the literature.  In Section
\ref{sec:further}
we illustrate with examples from the D-median on a line problem 
(placing web proxies on a linear topology network)
\cite{Wo00} and
wireless paging \cite{973511}.

\section{The Dynamic Programming Formulation}
\label{Sec:DP}

Set $S_0 = 0$ and $S_m= \sum_{i=1}^m p_i$ for $1 \le m \le n$.
Larmore and Przytycka \cite{larmore:1163}
formulated the binary Huffman coding problem as a DP (\ref{eq:LM})
where
$H(0) = 0$ and for $0 < i < n$:
\begin{equation}
\label{eq:LM}
H(i) = \min_{\max\{0,2i-n\} \le j < i} \left(H(j) + S_{2i-j}\right).
\end{equation}
In this DP, $H(n-1)$ is the cost of the optimal
Huffman code.  Another version of this DP, generalized for unequal-cost
binary coding alphabets, also appeared in \cite{BrGo02+}.

It is straightforward to modify
(\ref{eq:LM}) to model the binary LLHC problem.  The resulting DP is
\begin{equation}
\label{eq:DP_Def}
H(d,i) =
\left\{
\begin{array}{@{\,}ll}
0 & d = 0, i = 0 \\
\infty & d = 0, 0 < i < n \\
\displaystyle{\min_{0 \le j \le i} \left(H(d-1,j) + c^{(d)}_{i,j}\right)} & d > 0, 0 \le i < n
\end{array}
\right.
\end{equation}
where $H(D,n-1)$ will denote the cost of the optimal length-limited Huffman code and
\begin{equation}
\label{eqn:c}
c_{i,j}^{(d)} =
\left\{
\begin{array}{@{\,}ll}
0 & i = j = 0 \\
S_{2i-j} & \max\{0,2i-n\} \le j < i \\
\infty & \textrm{otherwise.}
\end{array}
\right.
\end{equation}
In the next subsection we will see an interpretation of this DP
(which also provides an interpretation of (\ref{eq:LM})).
In order to make this note self-contained, a complete derivation of the
DP for the $r$-ary alphabet case is provided in Appendix \ref{app:LLHCDP}.

As far as running time is concerned,
(\ref{eq:LM}) appears to a-priori require
$O(n^2)$ time to fill in its corresponding DP table.
\cite{larmore:1163} used the inherent concavity of $S_m$ to reduce this time
down to $O(n)$ by transforming the problem to an instance of
the
Concave Least Weight Subsequence (CLWS) problem and using one of the
known $O(n)$ time algorithms, e.g., \cite{Wi88}, for solving that problem.

Similarly, (\ref{eq:DP_Def}) appears to a-priori require
$\Theta(n^2D)$ time to fill in its DP table.  We will see that we may again
use the concavity of $S_m$ to reduce this down by an order of
magnitude, to $O(nD)$ by using the SMAWK algorithm \cite{AgKl87+}
for finding row-minima of matrices as a subroutine.
Unlike the CLWS algorithms, the SMAWK one is very simple to code and
very efficient implementations are available in different packages, e.g.,
\cite{Ep02,NeoBio}.  In the conclusion to this note, after the application
of the technique becomes understandable, we will explain
why \cite{larmore:1163} needed to use the more complicated CLWS
routine to solve the basic DP while we can use the simpler SMAWK one.

The $O(nD)$ DP algorithm for solving the LLCH problem,
while seemingly never explicitly stated in the literature, was known as folklore.
Even though it is much simpler to implement than the $O(nD)$
Larmore and Hirschberg \cite{LaHi90}
Package-Merge algorithm it suffers from the drawback of requiring $\Theta(nD)$ space.
The main contribution of this note is the observation that its space can be reduced down to
$O(n+D)$ making it comparable with Package-Merge.
Note that since, for the LLHC problem we may trivially assume $D \le n$, this implies
a space requirement of $O(n)$.
Furthermore, our
space improvement will work not only for the LLHC problem but for all
DPs in form (\ref{eq:DP_Def}) where the $c^{(d)}_{i,j}$ satisfy a particular property.

\subsection{The meaning of The DP}

\begin{figure*}[!htbp]
\centering
\includegraphics{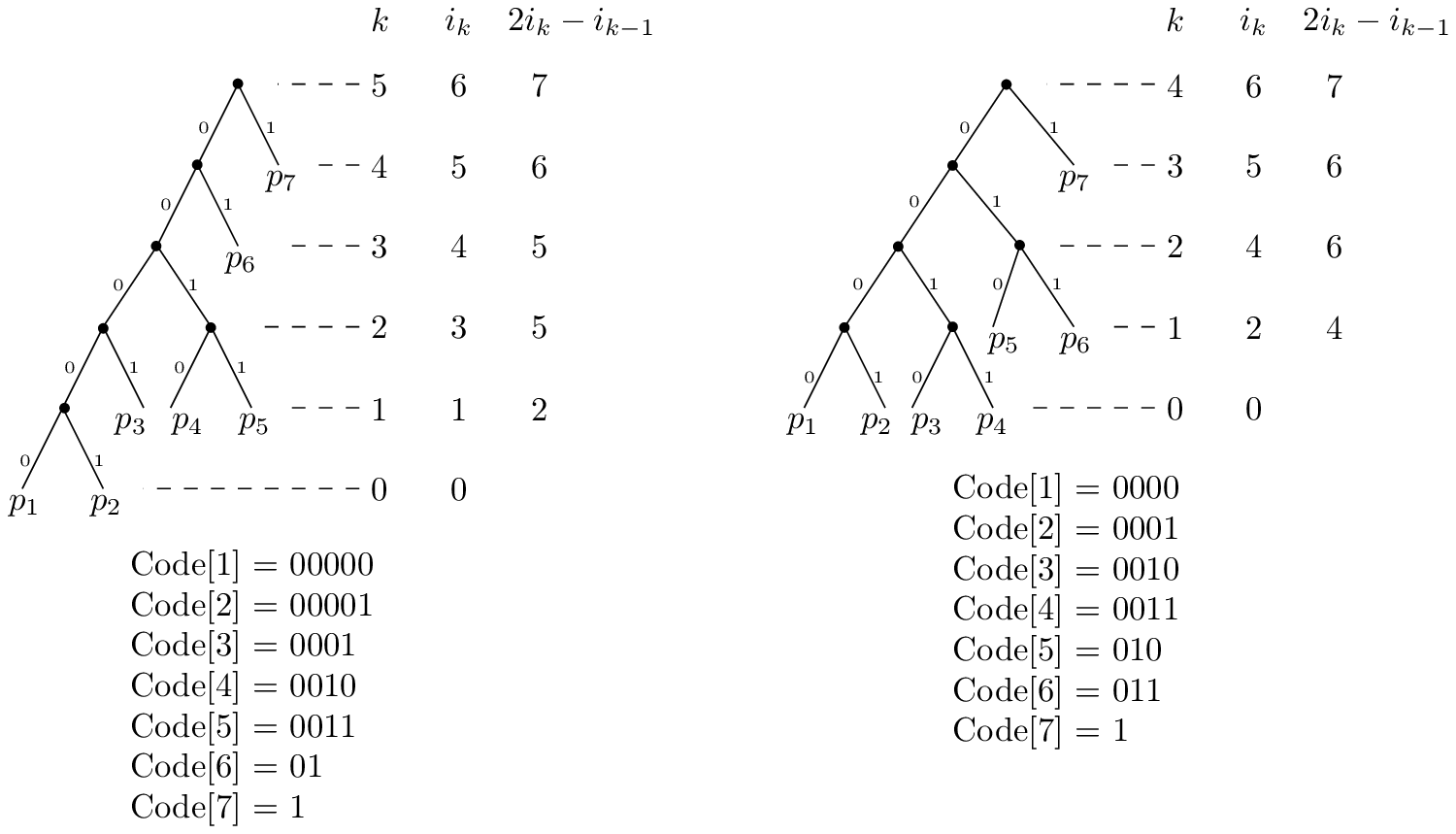}
\caption{Two trees and their corresponding sequences $\cali$ and codes.
The left tree has sequence $\cali_1 = (0,1,3,4,5,6).$ The right tree has
sequence $\cali_2 = (0,2,4,5,6).$ Note that, for both trees, $2 i_k - i_{k-1}$
is the number of leaves below level $k.$
For input frequencies $(p_1,\ldots,p_7) = (1,1,2,2,2,4,5,9)$.
The left tree is an  optimal Huffman code while
the right tree is an  optimal length-limited Huffman
code for $D = 4$.
Note that we allow padding sequences with initial {\bf 0}s, so the right tree could
also be represented by sequences $(0,0,2,4,5,6),$ $(0,0,0,2,4,5,6),$ etc.. 
}
\label{fig:seq_ex}
\end{figure*}

\begin{figure*}[!htbp]
\centering
\includegraphics{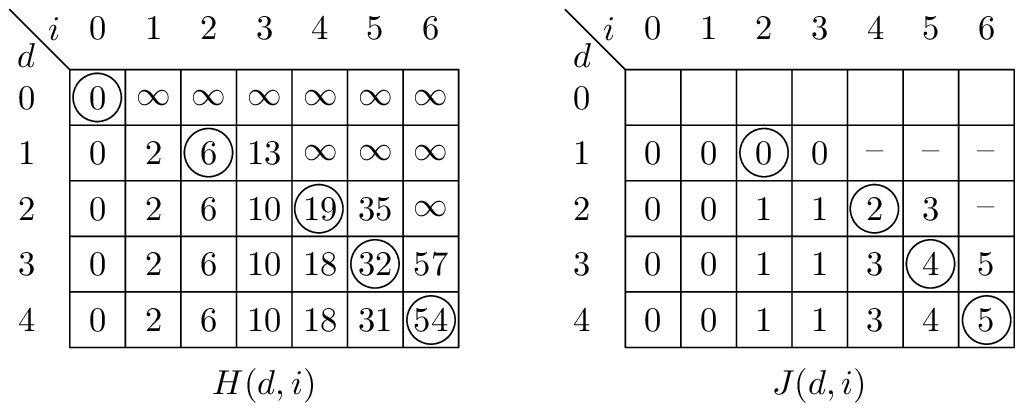}
\caption{Solving the DP in equation  \ref{eq:DP_Def} for
$(p_1,\ldots,p_7) = (1,1,2,2,2,4,5,9)$ with $D=4.$
$H(d,i)$ is the value defined by (\ref{eq:DP_Def});  $J(d,i)$
is the index $j$ for which the value $H(d,i)$ in (\ref{eq:DP_Def})
is achieved. The circled entries yield the sequence $(0,2,4,5,6)$  (the $6$
comes from the fact that we are calculating $H(4,{\bf 6})$) which
is exactly the sequence $\cali_2$  from  Figure
\ref{fig:seq_ex}.  The righthand tree in Figure
\ref{fig:seq_ex} is therefore an  optimal length-limited Huffman
code for $D = 4$. 
}
\label{fig:dp_table}
\end{figure*}

We quickly sketch the meaning of the DP (\ref{eq:DP_Def}) for the binary case.
Figures \ref{fig:seq_ex} and \ref{fig:dp_table} illustrate this sketch.
We note that in order to stress the parts important to our analysis,
our formalism is a bit
different than \cite{larmore:1163,BrGo02+}.  A complete derivation of the
DP for the $r$-ary case with the appropriate general versions
of the lemmas and observations stated below along with their proofs, 
is provided in Appendix \ref{app:LLHCDP}.

It is standard that there is a $1-1$ correspondence between
binary prefix-free code with $n$ words and binary tree with $n$
leaves.
The set of edges from an internal node to its children are labeled by a {\bf 0} or {\bf 1}.
Each leaf corresponds to a code word, which is the concatenation of the characters on the
root-to-leaf path.
The cost of the code equals the \emph{weighted external path length} of the tree. So
we are really interested in finding a binary tree with minimum weighted external path length.

Denote the height of the tree by $h.$
The bottommost leaves are on level $0$; the root on level $h$.
Optimal assignments of the $p_i$'s to the leaves always assign
smaller valued $p_i$'s to leaves at lower levels.

A node in a binary tree is \emph{complete} if it has two
children and  a tree is \emph{complete} if all of its internal nodes are complete.
A min-cost tree must be complete, so we restrict ourselves to complete trees.
A complete tree $T$ of height $h$ can be completely
represented by a sequence $(i_0, i_1, \ldots, i_h)$,
where $i_k$ denotes the number of internal nodes at levels $\le k$.  Note that,
by definition, $i_0 = 0$,  $i_h = n-1$.  Also note that every level must contain at least
one internal node so $i_0 < i_1 < \cdots < i_h$.  Finally,
it is straightforward (see Appendix \ref{app:LLHCDP}) to show that the total number of leaves
on level $< k$ is
$2 i_k - i_{k-1}$, so $2 i_k - i_{k-1} \le n$ for all $k$.  For technical reasons,
because we will be dealing with trees having height \emph{at most} $h$
(but not necessarily {\em equal to} $h$),
we allow initial padding of the sequence by
{\bf 0}s so a sequence representing a tree will be of the form
$(i_0, i_1, \ldots, i_h)$ that has the following properties


\begin{definition}
\label{def:valid.list}
Sequence $(i_0, i_1, \ldots, i_h)$ is \emph{valid} if
\begin{itemize}
\item $\exists t> 0 $ such that
$i_0 = i_1 = \cdots = i_t = 0$,
\item $0 < i_{t+1} <  i_{t+2} < \cdots < i_h \le n-1$
\item $2 i_k - i_{k-1} \le n$ for all
$1 \le k \le h$.
\end{itemize}
A sequence is \emph{complete} if it is valid and $i_h = n-1$.
\end{definition}

We can rewrite the cost function for a tree in terms of its complete sequence.
\begin{lemma}
\label{lem:eqn:cost.tree}
If complete sequence $(i_0, i_1, \ldots, i_h)$ represents a tree, then the cost of the tree is
$\sum_{k=1}^h S_{2  i_k - i_{k-1}}$.
\end{lemma}
(Note that padding  complete sequences with initial {\bf 0}s does not change the cost of the sequence.)

We may mechanically extend this cost function to {\em all}  valid
 sequences as follows.
\begin{definition}
\label{def:cost}
For valid  $\cali = (i_0, i_1, \ldots, i_h)$, set 
$$cost(\mathcal{I}) = \sum_{k=1}^h S_{2 i_k - i_{k-1}}.$$
$\cali$ is {\em optimal} if
$cost(\cali) = \min_{\cali'} cost(\cali')$ where the minimum is taken over all length $h$
sequences
$\cali' = (i'_0, i'_1, \ldots, i'_h)$ with $i'_h = i_h,$ i.e., all sequences of the same length that end with
the same value.
\end{definition}
Our goal is to find optimal trees by using the DP to optimize over valid sequences.  An
immediate issue is that not all complete sequences represent trees,  e.g.,
$\cali = (0,3,4,5)$ is complete for $n=6$ but,  by observation,  does not represent a tree.
The saving fact is  that even though not all complete sequences represent trees, all {\em optimal}
complete sequences represent trees.
\begin{lemma}
\label{lem:sequence.2.tree}
An optimal valid sequence ending in $i_h = n-1$ always represents a tree.
\end{lemma}

Thus,  to solve the LLHC problem of finding an
optimal tree of height $\le D$, we only  need to find an optimal valid sequence of length $h=D$
ending with $i_D=n-1$
(reconstructing the tree from the sequence can be done in $O(n)$ time).
In the  DP defined  by equations (\ref{eq:DP_Def}) and (\ref{eqn:c}), $H(d,j)$ clearly models
the recurrence for finding an optimal valid sequence $(i_0, i_1, \ldots, i_d)$ of length $d$ with $i_d=j$ so
this DP  solves the problem.

Note that, a-priori,   filling in the DP table $H(\cdot,\cdot)$ one entry at a time seems to  require
$O(n^2D)$ time.  We will now sketch the standard way of reducing this time down to $O(nD).$  Before doing
so we must  distinguish between the {\em value} problem and the {\em construction problem}.  The value
problem would be to calculate the value of $H(D,n-1).$  The construction problem would be to
construct an optimal valid {\em sequence}
$\cali = (I_1,I_2,\ldots,I_{D})$  with $I_{D}=n-1$
and  $cost(\cali) = H(D,n-1).$
This would require  backtracking through the DP table by
setting $I_0=0$, $I_{D} = n-1$ and finding  $I_1, I_2, \ldots I_{D-1}$ such that
\begin{equation}
\label{eq:const}
\forall 0 < d \le D,\quad H(d,I_d) = H(d-1,I_{d-1}) + c^{(d)}_{I_d,I_{d-1}}.
\end{equation}

\subsection{Solving the Value problem in $O(nD)$ time}

\begin{figure*}[!htbp]
\centering
\includegraphics{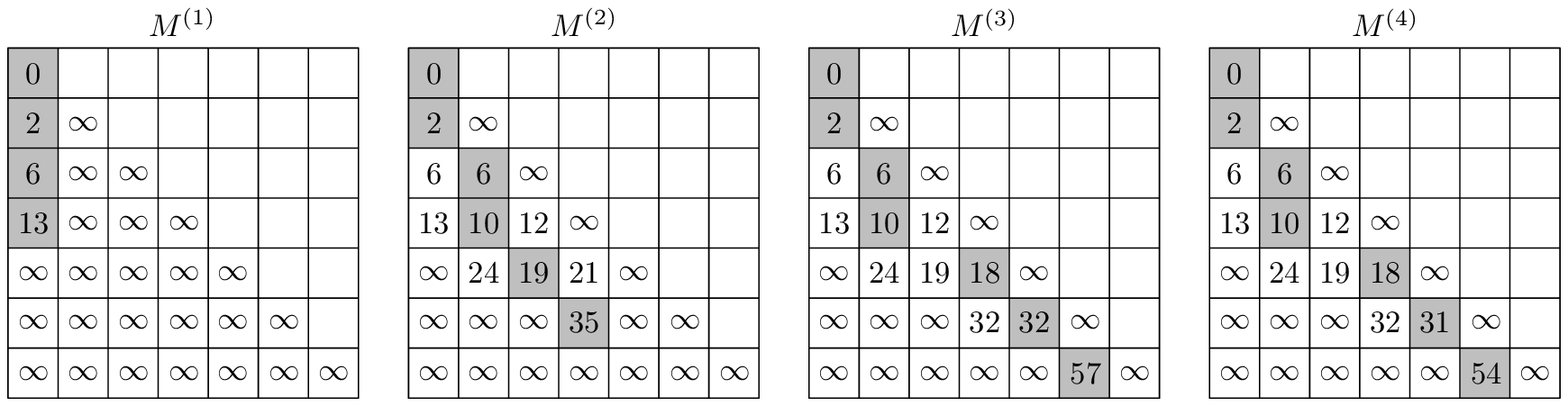}
\caption{The matrices used for calculating the DP tables in Fig.~\ref{fig:dp_table}.
The shaded entries are the row minima. The row minima for $M^{(i)}$ are exactly the
row entries in the $H(d,i)$ table in Fig.~\ref{fig:dp_table}. The column indices of
the corresponding row minima are the $J(d,i)$ entries.}
\label{fig:row_ex}
\end{figure*}

\begin{definition}
An $n \times m$ matrix $M$ is
{\em Monge}\footnote{This property is sometimes alternatively defined by:
for  $0 \le i < i' < n$ and $0 \le j < j' <  m$
$M_{i,j} + M_{i',j'} \le M_{i',j} + M_{i,j'}$ but it is well known,
see, e.g., \cite{BuKl96+}, that this is equivalent to (\ref{eq:Monge def}).}
if for $0 \le i < n-1$ and $0 \le j <  m-1$
\begin{equation}
\label{eq:Monge def}
M_{i,j} + M_{i+1,j+1} \le M_{i+1,j} + M_{i,j+1}
\end{equation}
\end{definition}
The Monge property can be thought of as a discrete version of concavity.  It appears implicitly
in many optimization problems for  which it permits speeding up their solutions (\cite{BuKl96+})
provides a nice survey).
One of the classic techniques used is the SMAWK algorithm for finding row-minima.

Given an $n \times m$ matrix $M$, the {\em minimum} of row $i$, $i=1,\ldots,n$ is the entry of row $i$ that
has the smallest value; in case of ties, we take the rightmost entry.
Thus, a solution of the row-minima problem is a collection of indices $j(i)$, $i=1,\ldots,n$ such
that
{\small
\begin{equation}
\label{eq:res}
M_{i,j(i)} = \min_{0 \le j < m} M_{i,j}
\mbox{ and }
\hspace*{.02in}
j(i) = \max \{j \,:\, M_{i,j} = M_{i,j(i)}\}.
\end{equation}
}
Figure \ref{fig:row_ex} gives four examples of Monge matrices and their row minima.

At first glance it seems that we would have to examine all of the $mn$ entries in $M$ to find the
row minima but, \cite{AgKl87+} proved\footnote{Technically,
\cite{AgKl87+} proved their result for a larger class, the {\em totally-monotone matrices}.  But all
applications in the literature seem to be for Monge matrices.}
\begin{lemma} (The SMAWK algorithm \cite{AgKl87+})\\
\label{lem:SMAWK}
Let $M$ be a $n \times m$ Monge matrix such that entry $M_{i,j}$ can be calculated in
$O(1)$ time.
Then the row minima problem on $M$ can be solved in  $O(n + m)$ time.
\end{lemma}

The constant hidden by the $O(\,)$ is very small, around 2, and the algorithm is easy to code,  so it
is quite practical to use.

Note that the SMAWK algorithm doesn't have the time available to build the entire $n \times m$
 matrix.  Instead,  it searches through the matrix in a clever way,  constructing entries as needed.
One standard use of the SMAWK algorithm is  in the speedup of dynamic programs that have Monge properties.

\begin{definition}
A  DP in the form (\ref{eq:DP_Def}) is  {\em Monge} if,
for all $1 \le d \le D$ and $0 \le j \le i < n$,
\begin{equation}
\label{eq:def_Quad}
c^{(d)}_{i,j} + c^{(d)}_{i+1,j+1} \le c^{(d)}_{i+1,j} + c^{(d)}_{i,j+1}
\end{equation}
\end{definition}

{\small \em Note: In many DP applications, it is possible that for some $i,j$,
$c^{(d)}_{i,j} = \infty$.  The inequality in (\ref{eq:def_Quad}) treats
$\infty$ in the natural way, e.g., for any constant $c$;  $c < \infty$ and   $c + \infty = \infty$. Also,
$\infty + \infty = \infty$. The SMAWK algorithm permits the use of $\infty$ in this way.}

Now suppose that a DP  defined by (\ref{eq:DP_Def}) is Monge.
For $d=1,2,\ldots,D$ define matrix $M^{(d)}$ by
$$
M^{(d)}_{i,j} =
\left\{
\begin{array}{ll}
H(d-1,j) + c^{(d)}_{i,j} & \mbox{if $0 \le j \le i < n$} \\
\infty & \mbox{otherwise}
\end{array}
\right.
$$
Then, from (\ref{eq:def_Quad}), we have
{\footnotesize
\begin{eqnarray*}
\hspace*{-.15in} M^{(d)}_{i,j} + M^{(d)}_{i+1,j+1}
&=& H(d-1,j) + H(d-1,j+1) + c^{(d)}_{i,j} +c^{(d)}_{i+1,j+1}\\
&\le& H(d-1,j) + H(d-1,j+1) + c^{(d)}_{i+1,j} +c^{(d)}_{i,j+1}\\
&=& M^{(d)}_{i+1,j} + M^{(d)}_{i,j+1}
\end{eqnarray*}
}
and  $M^{(d)}$ is Monge.
Note that
\begin{eqnarray*}
H(d,i) &=&  \min_{0 \le j \le i} \left(H(d-1,j) + c^{(d)}_{i,j}\right)\\
&=& \min_{0 \le j \le i} M^{(d)}_{i,j}  =  \min_{0 \le j \le N} M^{(d)}_{i,j}.
\end{eqnarray*}
So, $H(d,i)$ are just the row-minima of $M^{(d)}$.
See Figure \ref{fig:row_ex}.
Since $M^{(d)}$ is Monge, we can use the SMAWK algorithm to, in $O(n)$ time, find {\em all} of its
row minima at one time.
More specifically,  let $J(d,i)$ and
$M^{(d)}_{i,J(d,i)}$ be the corresponding
values (\ref{eq:res}) returned when running SMAWK$(M^{(d)}).$
Then the algorithm for filling in the table is just to iteratively
run down the rows of the table, using SMAWK to fill in each  row by using knowledge of
the previous row:

\begin{figure}[h]
\par\noindent\underline{\tt Fill\_Table}\\
For $d=1$ to $D-1$\\
\hspace*{.1in} SMAWK $(M^{(d)})$\\
\hspace*{.1in} $\forall 0 \le i < n$ set   $H(d,i) = M^{(d)}_{i,J(d,i)}$
\caption{The $O(nD)$ algorithm for the value problem.}
\label{fig:alg_FT}
\end{figure}


Note that this algorithm uses $\Theta(nD)$ time, since, for each fixed $d$, the SMAWK algorithm
only uses $O(n)$ time.
Also note that if we're only interested in the final row, then the algorithm uses only
$O(n)$ space, since once row $d$ has been calculated,  the values from row $d-1$ can be thrown away.


We now return to the LLHC problem and show that it can be plugged into the above machinery.
\begin{lemma}
\label{lem:c.qi}
The $c_{i,j}^{(d)}$ defined in (\ref{eqn:c}) satisfy Monge property
(\ref{eq:def_Quad}).
\end{lemma}
\begin{IEEEproof}
If $i=j=0$ the righthand side of (\ref{eq:def_Quad}) is $\infty$, so
(\ref{eq:def_Quad}) is satisfied.

If $j+1 = i$ or $2(i+1)-n > j$, the righthand side of (\ref{eq:def_Quad}) is $\infty$, so
(\ref{eq:def_Quad}) is satisfied.

If $j+1 < i$ and $2(i+1)-n \le j$, (\ref{eq:def_Quad}) can be rewritten as
\begin{equation}
\label{eqn:c.qi.rewrite}
S_{2i-j} + S_{2(i+1)-(j+1)} \le  S_{2i-(j+1)} + S_{2(i+1)-j}
\end{equation}
It is easy to verify
\begin{eqnarray*}
S_{2i-j} + S_{2(i+1)-(j+1)} - S_{2i-(j+1)} - S_{2(i+1)-j}&&\\ 
\hspace*{.2in} = p_{2i-j} - p_{2i-j+2} \le 0&&
\end{eqnarray*}

Hence, (\ref{eqn:c.qi.rewrite}) holds.
\end{IEEEproof}

Thus,  from the discussion above,  we can find all of the $H(d,i)$ in $\Theta(nD)$ time.
In particular, $H(D,n-1)$ will be the cost of the optimal tree with height {\em at most} $D$
which is the required cost of the optimum $D$-limited code.

We have thus seen how to solve the value problem in $O(nD)$ time.
The difficulty is that {\em constructing}  the optimal tree associated with
$H(D,n-1)$  would require
finding the associated optimal valid sequence with $i_{D} = n-1.$  This would require solving
the construction problem by finding all indices $I_d$ in
(\ref{eq:const}).  The standard way of  solving this problem is to
 maintain  an   array storing the
$J(d,i)$ values returned by the algorithm.
%
%
Starting from   $H(D,n-1)$ and backtrack through the $j(\cdot,\cdot)$ array, constructing the
corresponding sequence by setting $I_{D} = n-1$ and $I_{d-1} = j(d,I_d).$
Unfortunately, this requires maintaining a size $\Theta(nD)$ auxiliary array,  which
requires too much space.

\section{ Solving the Construction Problem in $O(nD)$ time and $O(n+D)$ space}
\label{Sec:space}

\begin{figure}[!htbp]
\centering
\includegraphics{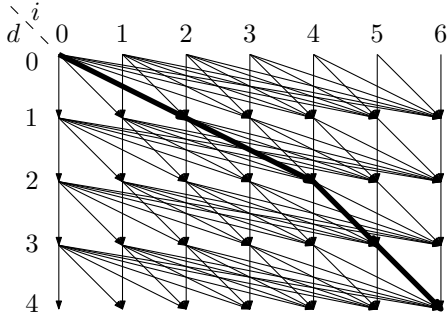}
\caption{The dropping-level graph associated with the example from Figures
\ref{fig:dp_table} and  \ref{fig:row_ex}.  The bold edges are the
minimum cost path from $(0,0)$ to $(6,4).$  Note that the $i$ coordinates
of the path are $(0,2,4,5,6)$ which is {\em exactly} the sequence of $J(d,i)$'s corresponding to optimal solution of the problem,  which is also the sequence
corresponding to the optimal tree.
}
\label{fig:drop_ex}
\end{figure}

Let $V$ be the grid nodes $(d,i)$ with $0 \le d \le D$ and $0 \le i <n$.
Consider the directed graph $G=(V,E)$ in which $(d,i)$ points  to all nodes immediately below it and to its  right, i.e.,
$$E = \{(\, (d,j),\, (d+1,i)\,) \mid  (d,j) \in V,\,  d < D,\,  j\le i \}$$
See Figure \ref{fig:drop_ex}.
Such graphs are sometimes called {\em dropping level-graphs} \cite{MuRa82}.
Now assign edge $(\, (d-1,j),\, (d,i)\,)$ the weight $c^{(d)}_{i,j}$.
The length of a path in $G$ will just be the sum of the weights of the edges in the path.
The important observation is that $H(d,i)$ in  DP (\ref{eq:DP_Def}) is simply
the length of the min-cost path from $(0,0)$ to $(d,i)$ in this weighted $G$. More
specifically,  the value
problem is to find the {\em length of a shortest path} and the construction problem is
to find an  {\em actual shortest path.}

A-priori,  finding such  a path seems to  require $O(nD)$ space.  There are two different
algorithms in the literature for reducing the space down to $O(n+D)$ in  related problems.

The first was for finding a 
maximum common subsequence of two sequences. This reduced down to the problem of finding
a {\em max-length} path in something very similar to a dropping level-graph in which each vertex has
bounded indegree and bounded outdegree.  Hirschberg \cite{Hi75} developed an $\Theta(nD)$ time, $\Theta(n+D)$ space algorithm for this problem.  His algorithm was very influential in the bioinformatics community and its technique is incorporated into many later algorithms e.,g \cite{ChHa94+,MyMi88}. The techniques's performance  is very dependent upon the bounded degree of the vertices,  which is not true in our case.

The second, due to  Munro and Ramirez \cite{MuRa82},  was exactly for the problem of constructing
min-cost  paths in full dropping level-graphs.  Their algorithm ran in $\Theta(n^2D)$ time and $\Theta(n+D)$ space.
Their $\Theta(n^2D)$ time is too expensive for us.  We will now see how to reduce this down to
$\Theta(nD)$ using the Monge speedup while still maintaining the $\Theta(n+D)$ space.

The general problem  will be
to construct an optimal $u$-$w$ path in $G$ where $u=(d_u,i_u)$ is above and not to the left of  $w=(d_w,i_w),$
i.e., $d_u < d_w$ and $i_u \le i_w.$
Let $G(u,w)$ be the subgrid with upper-left corner  $u$ and lower-right corner
$w$ (with associated induced edges from $G$).
First note that, because $G$ is a dropping level-graph, 
any optimal (min or max cost) $u$-$w$ path in $G$ must lie completely in $G(u,w)$.
Both algorithms  \cite{Hi75,MuRa82} start  from the same observation, which is  to  build the
path {\em recursively} i.e.,   by first (a)
 finding a point $v=(\bar d, \bar i)$ halfway (by link distance) on the optimal $u$-$w$ path
 in $G(u,w)$ and then (b) output the recursively constructed optimal $u$-$v$ path  in $G(u,v)$ and
 optimal $v$-$w$ path  in $G(v,w)$.

For dropping level-graphs, if $u=(d_1,i_1)$ and $w=(d_2,i_2)$ then the midlevel must be
$\bar d = \lfloor {(d_1 + d_2)}/2 \rfloor $.  Suppose that we had an algorithm $Mid(u,w)$ that returned
a  point $v=(\bar d, \bar i)$ on a shortest $u$-$w$ path in $G(u,w).$  Then, translated into our notation and
with appropriate termination conditions
 the construction algorithm can be written as:
 
\vspace*{.5in}
 
 \begin{figure}[h]
 \par\noindent\underline{{\tt Path}$(u,w)$}\\
 1. If $u=(d,j)$ and $w=(d+1,i)$ then\\
 2. \hspace*{.1in} output edge $(u,w)$\\
 3. Else if $u=(d,i)$ and $w=(d',i)$ then\\
 4.  \hspace*{.1in} Output  vertical path from $u$ to $w$\\
 5. Else\\
 6. \hspace*{.1in}  set $v=Mid(u,w)$\\
 7. \hspace*{.1in} {\tt Path}$(u,v)$; \ {\tt Path}$(v,w)$\\
  \label{fig:alg_Path}
  \caption{The algorithm for constructing a min-cost $u$-$w$ path.}
  \end{figure}

\begin{figure*}[!htbp]
\centering
\includegraphics{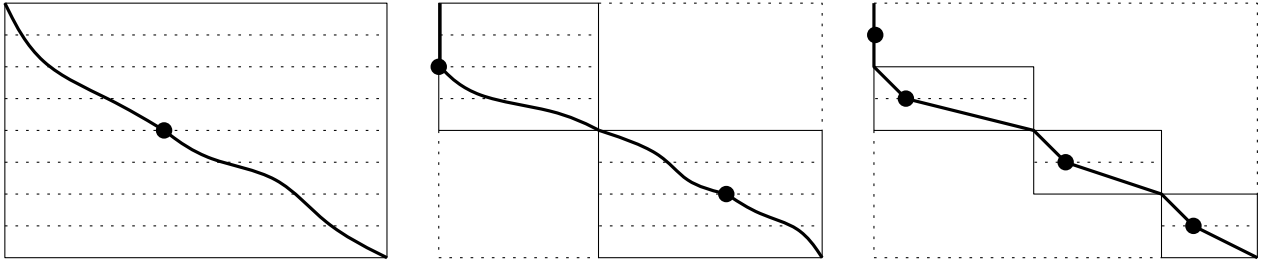}
\caption{An illustration for finding the optimal path.
Here, $D = 8$ and there are $3$ levels of recursions.
The solid circles are the intermediate nodes found by the $Mid(u,v)$ procedures.
The first level of recursion finds the midpoint on level $4;$ the second level, the midpoints on
levels $2$ and $6$; the third the midpoints on levels $1,3,5,7.$  At that point all
subproblems are of height one and easily solvable. Note that each recursive call splits
a problem on a box of height $2^i$ into two problems on disjoint boxes of height $2^{i-1}.$}
\label{fig:path_illus}
\end{figure*}

(Figure \ref{fig:path_illus} illustrates this idea.)
 To solve the original problem we just call {\tt Path}$(u_0,w_0)$
 where $u_0=(0,0)$ and $w_0=(D,n-1).$
 Correctness follows from the fact that at each recursive call,
 the vertical distance $d_w-d_u$ decreases so the recursion must terminate.
 Furthermore, when the recursion terminates,  either (i)
 $u=(d,j)$ and $w=(d+1,j)$ so  the {\em only} $u$-$w$
 path in $G(u,w)$  is the edge $(u,w)$ or (ii) $u=(d,i)$ and $w=(d',i)$
 so  the {\em only} $u$-$w$
 path in $G(u,w)$  is the vertical path going down from $u$ to $w$.

 The efficiency of the resulting algorithm, both in time and space,  will depend upon how efficiently  $v=Mid(u,w)$ can be found.
 Note that with the exception of the calls of type 
  $Mid(u,w)$, the  rest of the execution   of {\tt Path}$(u_0,w_0)$ (including all recursive calls) 
 only requires  a total of $O(D)$ space,  since each recursive call  uses
 only $O(1)$ space and there are at most $O(D)$ such calls.  Thus,  if  $Mid(u,w)$ can be
  found  using $O(n+D)$ space,  then the entire procedure requires only $O(n+D)$ space.
  This is actually how both
 \cite{Hi75,MuRa82} achieve their  space bounds.  The two algorithms differ in how they calculate $v$.
   Although both their approaches  can be used  for our problem,  we will work with a modified version of that of \cite{MuRa82}, since it will be simpler to explain.

 We now describe how to use the SMAWK algorithm to find
 $Mid(u_0,w_0)$ in $O(nD)$ time and $O(n)$ space.  The  extension to general $Mid(u,w)$
 will follow later.  Recall that the procedure {\tt Fill\_Table} from Figure  \ref{fig:alg_FT}
  used the  fact that
 $H(\cdot,\cdot)$ was Monge and  the SMAWK algorithm
  to iteratively fill in the rows $H(d,\cdot)$,  for $d=1,2,\ldots,D.$  Given row $H(d-1,\cdot)$,
 the  procedure
 calculated $H(d,\cdot)$ in $O(n)$ time using SMAWK,   and then threw away $H(d-1,\cdot)$.

Consider an arbitrary node $(d,i)$ on level $d > \bar d.$
The shortest path from $u_0$ to $(d,i)$ must pass through {\em some} node on level $\bar d$.
We now modify {\tt Fill\_Table}
 to ``remember''  this  node.
More specifically, our algorithm will  calculate  auxiliary data  $pred(d,i)$.
\begin{itemize}
\item For $d < \bar d$, $pred(d,i)$ will be undefined.
\item For $d \ge \bar d$, $pred(d,i)$ will be an
index $j$ such that node $(\bar d,j)$  appears
on some  shortest path from $u_0$ to $(d,i)$.
\end{itemize}

So, when the procedure terminates,  $v=(\bar d,\,  pred(d, n-1))$ will be $Mid(u_0,w_0).$

By definition, on level $\bar d$,  we have  $pred(\bar d,i) =i.$

For $d > \bar d$ suppose $(d-1,j')$ is the immediate predecessor of $(d,i)$ on the shortest path
from $u_0$ to $(d,i)$.  Then (i) a  shortest path from $u_0$ to $(d-1,j')$ followed by 
(ii) the edge from
$(d-1,j')$ to 
$(d,i)$ is  (iii) a shortest
path from $u_0$ to  $(d,i)$; we may therefore set $pred(d,i) = pred(d-1,j').$

We can use this observation to  modify {\tt Fill\_Table}
to calculate  the $pred(d,\cdot)$ information.

\begin{figure}[h]
\par \noindent \underline{$Mid(u_0,w_0)$}\\
For $d=1$ to $\bar d$\\
\hspace*{.1in} SMAWK $(M^{(d)})$\\
\hspace*{.1in} $\forall 0 \le i < n$ set   $H(d,i) = M^{(d)}_{i,J(d,i)}$\\
$\forall 0 \le i < n$ set $pred(\bar d,i) = i;$\\
For $d=\bar d +1 $ to $D$\\
\hspace*{.1in} SMAWK $(M^{(d)})$\\
\hspace*{.1in} $\forall 0 \le i < n$,\ set   $H(d,i) = M^{(d)}_{i,J(d,i)}$\\
\hspace*{.1in} $\forall 0 \le i < n$,\  set $pred(d,i)  = pred\left(d-1,j(d,i)\right)$
\caption{Returns the midpoint, by link distance, on min-cost $u_0$-$w_0$ path.}
\label{fig:Midalg}
\end{figure}
  Note that $Mid(u_0,w_0)$ can throw away all
of the values $pred(d-1,\cdot)$ and $H(d-1,\cdot)$ after the values $pred(d,\cdot)$ and $H(d,\cdot)$ have been calculated, so it only uses  $O(n)$ space. Similarly to the analysis of
{\tt Fill\_Table}, it uses only $O(nD)$ time since each call to the SMAWK algorithm uses only
$O(n)$ time.

 So far,  we have only shown how to find $v=Mid(u_0,w_0)$.
 Note that the {\em only assumptions} we used were that $H(\cdot,\cdot)$ satisfies
 DP (\ref{eq:DP_Def}) and  is Monge, i.e.,
 the $c^{(d)}_{i,j}$ satisfy  (\ref{eq:def_Quad}).

Now suppose that we are given 
$$u=(d_u,i_u),\   w=(d_w,i_w) \mbox{ with $d_u < d_w$ and $i_u \le i_w$}.$$
$G(u,w)$ is  a dropping level-graph on its own nodes so the cost of the shortest path from $u$ to any node $(d_u+d,i_u +i) \in G(u,w)$ is $\tilde H(d,i)$
defined by
{\small
\begin{equation}
\label{eq:DP_Def_2}
\tilde H(d,i) =
\left\{
\begin{array}{ll}
0 & \mbox{if $d = 0$, $i = 0$} \\
\infty & \mbox{if $d = 0$, $0 < i <N$} \\
\displaystyle{\min_{0 \le j \le i} \left(\tilde H(d-1,j) + \tilde c^{(d)}_{i,j}\right)} & \mbox{if $d>0$, $0 < N$}
\end{array}
\right.
\end{equation}
}
where $N=i_w-i_u +1$ and
$\tilde c^{(d)}_{i,j} = c^{(d)}_{i_u+ i,i_u + j}.$
Note that this new DP is exactly in the same form as (\ref{eq:DP_Def}),  just with a different $n$
and shifted $c^{(d)}_{i,j}.$
Since the original $c^{(d)}_{i,j}$ satisfy (\ref{eq:def_Quad}),
so do the $\tilde c^{(d)}_{i,j}$.  Thus (\ref{eq:DP_Def_2}) with the
$\tilde c^{(d)}_{i,j}$ is Monge as well.

Therefore,  we can
run exactly the same algorithm written in Figure \ref{fig:Midalg}  to find
the midpoint
 $v=(\bar d, \bar i)= Mid(u,w),$
 of the min-cost $u$-$w$ path in $O((d_w -d_u)N)$ time and
 $O((d_w -d_u)+ N)= O(D+n)$ space.

 As discussed  previously,  if $Mid(u,w)$ only requires $O(n+D)$ space, then  {\tt Path}$(u,w)$ only
 requires $O(n+D)$ space, so we have completed the space analysis.

 It remains to analyze running  time. Set
 $$Area(u,w) = (N-1) (d_w-d_u)
 $$
 to be the ``area'' of $G(u,w)$.
 Recall that line 3 of  {\tt Path}$(u,w)$implies  that $d_u \not = d_w$ when $Mid(u,w)$ is called.
 Therefore  $N \ge 1$ and the running time of
 $Mid(u,w)$ is
 $$O((d_w -d_u)N) = O(Area(u,w)).$$

We now analyze the {running time} of {\tt Path}$(u_0,w_0)$.
First consider the recursive calls
 when lines 1-4  occur, i.e., the recursion terminates.
 The {\em total} work performed by such calls is the total number of
 edges outputted.  Since an edge is outputted only once and
 the total path contains $D$ edges, the total work performed is $O(D).$

Next consider the calls when line 5-7  occur.  Since each such call returns a vertex $v$ on the path,
 there are only $D-1$ such calls so lines 6 and 7 are only called $O(D)$ times and their total work, with
 the exception of the call to $Mid(u,v)$, is $O(D).$

Finally consider the work performed by the $Mid(u,w)$ calls.  Partition the calls into levels.
\begin{itemize}
\item  Level 1 is the original call $Mid(u_0,w_0)$.
\item Level 2 contains the recursive calls directly
made by the level-1 call.
\item In general,  level $i$ contains the recursive calls  directly
made by the level-$(i-1)$ calls.
\end{itemize}
Note that if  $Mid(u,w)$ is a level $i$ call
with $u=(d_u,i_u)$ and $w=(d_w,i_w)$) then 
$$\frac D {2^i} \le d_w - d_u <  \frac D {2^i} + 1.$$
Furthermore,  by induction,
if  $Mid(u,w)$ and $Mid(u',w')$ are two different  level $i$ calls, then horizontal ranges
$[d_u,d_w]$ and $[d_{u'},d_{w'}]$ are
{\em disjoint} except for possibly $d_w = d_{u'}$ or $d_u = d_{w'}$.

Fix $i.$  Let $(u_j,w_j)$ $j=1,\ldots t$ be the calls at level $i$.  The facts that each grid $G(u_j,w_j)$
has height $\le \frac D {2^i} + 1$ and that the horizontal ranges of the grids are disjoint implies
$$\sum_{j=1}^t Area(u_j,w_j) \le n \left(\frac D {2^i}+ 1\right).$$

Thus the total
of all level-$i$ calls is $O\left( n \left(\frac D {2^i} + 1\right)\right)$. Summing over the $\lceil \log D \rceil$
levels we get  that the total work performed by all of the $Mid(u,w)$ calls on line 6 is
$$O\left( \sum_i   n \left(\frac D {2^i} + 1\right)\right) = O(nD).$$
Thus,  the total work performed by {\tt Path}$(u_0,w_0)$ is $O(nD)$ and we are finished.

\section{Further Applications}
\label{sec:further}

We just saw how, in $\Theta(nD)$ time and $\Theta(n+D)$ space,
to solve the construction problem for any  DP in form  (\ref{eq:DP_Def})
that satisfies the Monge property (\ref{eq:def_Quad}).
$\Theta(nD)$ time was known previously;  the $\Theta(n+D)$ space bound, is
the new improvement.
There are many other DP problems
besides the binary LLHC that satisfy (\ref{eq:def_Quad}) and whose space
can thus be improved.
We illustrate with three examples.

\medskip

\par\noindent\underline{The $r$-ary LLHC problem:}\\
We have discussed the binary LLHC problem in which $|\Sigma|=2.$  The general $r$-ary alphabet case
with $N$ probabilities  is
still modeled by a DP in form  (\ref{eq:DP_Def}) but with $n = \frac {N-1} {r-1} + 1$.
 The only difference is that (\ref{eqn:c})
is replaced by
\begin{equation}
\label{eqn:c_r}
c_{i,j}^{(d)} =
\left\{
\begin{array}{ll}
S_{ri-j} & \textrm{if } \max\{0,ri-N\} \le j < i \\
\infty & \textrm{otherwise.}
\end{array}
\right.
\end{equation}
A full derivation of this DP is given in Appendix \ref{app:LLHCDP}.
The proof that the
$c_{i,j}^{(d)}$ satisfy the Monge property (\ref{eq:def_Quad}) is similar to  the proof of
Lemma \ref{lem:c.qi}.
Thus,  we can construct a solution to the $r$-ary LLHC problem in $\Theta(ND)$ time and $\Theta(N)$ space as well.

\medskip

\par\noindent\underline{$D$ medians on a line:}\\
We are given $n-1$ customers located on the positive real  line;  customer $i$ is at location $v_i.$
Without loss of generality, assume $v_1 < v_2 < \cdots < v_{n-1}.$  There are $D \le n$ service centers located
on the line and a customer is serviced by the closest service center to its left (thus we always assume
a service center at $v_0=0$).  Each customer has a service request $w_i >0.$  The cost of servicing
customer $i$ is $w_i$ times the distance to its service center.  In \cite{Wo00}, motivated by
the application of optimally placing web proxies on a linear topology network,
 Woeginger showed that this
problem could be modeled by a DP in form in form  (\ref{eq:DP_Def}) where
$$c^{(d)}_{i,j} = \sum_{l=j+1}^i w_l(v_l-v_{j+1})$$ 
and proved that these $c^{(d)}_{i,j}$ satisfy Monge property
(\ref{eq:def_Quad}).  He then used the SMAWK algorithm to construct a solution in $O(nD)$ time and $O(nD)$ space.
Using the technique we just described,  this can be reduced to $O(nD)$ time and $O(n)$ space.

We also mention that there is an undirected variant of this problem in which a node is serviced by its {\em closest}
service center looking both left and right.  There are many algorithms in the literature that
(explicitly or implicitly) use concavity to construct solutions for this problem
 in $O(nD)$ time using $O(n)$ space,
e.g., \cite{HaTa91,AuPa98+,FGZ06}. \cite{FGZ06} does this by using a DP formulation that is in the
DP form (\ref{eq:DP_Def})
and satisfies the Monge property (\ref{eq:def_Quad}) so
the technique in this paper can reduce the space  for this problem down to
$O(n)$ as well.

\medskip

\par\noindent\underline{Wireless Paging:}\\
The third application comes from wireless mobile paging. A user can be in one of $N$
different cells. We are given   
a probability distribution in which $p_i$ denotes the probability that a  
user will be in cell $i$ and want to minimize the bandwidth needed to send paging requests to
identify the cell in which the user resides.
This problem was originally conjectured to be NP-complete, but \cite{KGW04+}  developed a
DP algorithm for it.
The input of the problem is the $n$ probabilities $p_1 \ge p_2 \ge \cdots \ge p_n$ and an integer
$D\le n$ (corresponding to the number of paging rounds used).
The DP developed by \cite{KGW04+} is exactly in our DP  form (\ref{eq:DP_Def})
with
\begin{equation}
\label{eq:wmp.cd}
c^{(d)}_{i,j} = \left\{
\begin{array}{ll}
i \left(\sum_{\ell=j+1}^i p_{\ell}\right) & \textrm{if } d-1 \le j < i \\
\infty & \textrm{otherwise.}
\end{array}
\right.
\end{equation}
The goal is to compute $H(D,n)$, which will be the minimum expected bandwidth needed.
Solving the construction version of this DP  permits constructing the actual paging protocol that yields this minimum
bandwidth.

\cite{KGW04+} used the naive algorithm to solve the DP in $\Theta(n^2D)$ time and $\Theta(nD)$ space.
\cite{BGZ06} proved that the $c^{(d)}_{i,j}$ defined by (\ref{eq:wmp.cd}) satisfy the
the Monge property (\ref{eq:def_Quad}) and thus reduced the time to $\Theta(nD)$, but still
required  $\Theta(nD)$ space.
The algorithm in this paper permits improving the space complexity of constructing the protocol
down to $\Theta(n)$.

\section{Conclusion}
The standard approach to  solving the Length-Limited Huffman Coding (LLHC) problem is via the
special purpose Package-Merge algorithm of  Hirschberg and Larmore \cite{LaHi90} which runs in $O(nD)$
time and $O(n)$  space, where $n$ is the number of codewords and $D$ is the length-limit on the code.

In this note we point out that this problem can be solved in the same time and space using a straightforward
Dynamic Programming formulation.  We started by noting that it was known that
the LLHC problem could  be modeled using a DP in the form
{\small
\begin{equation}
\label{eq:H_final}
H(d,i) =
\left\{
\begin{array}{ll}
0 & \mbox{if $d = 0$, $i = 0$} \\
\infty & \mbox{if $d = 0$, $0 < i <n$} \\
\displaystyle{\min_{0 \le j \le i} \left(H(d-1,j) + c^{(d)}_{i,j}\right)} & \mbox{if $d>0$, $0 < i < n$}
\end{array}
\right.
\end{equation}
}

where $H(d,n)$ will denote the minimum cost of a code with longest word at most $d$ and the
$c^{(d)}_{i,j}$ are easily calculable constants. This implies an $O(n^2D)$ time $O(nD)$ space algorithm.
We then note that, using standard DP speedup techniques, e.g., the SMAWK algorithm,  the time could be
reduced down to $O(nD)$.  The main contribution of this paper is to note that, once the problem is
expressed in this formulation,   the space can be reduced down to $O(n)$ while maintaining the
time at $O(nD)$.  The space reduction developed for this problem was also shown to apply to other problems
in the literature that previously had been thought to require $\Theta(nD)$ space.

We conclude by noting that if we're only interested in solving the standard Huffman coding problem and not the LLHC one then DP (\ref{eq:H_final}) with
$c^{(d)}_{i,j}$ defined by (\ref{eqn:c_r}) collapses down to
\begin{equation}
H(i) = \min_{\max\{0,ri-N\} \le j < i} H(j) + S_{ri-j}.
\end{equation}
where $H(i)$ denotes the minimum cost of a ``valid sequence'' ending in $i$.
$H\left(\frac  {N-1} {r-1}\right)$ will be the cost 
of an optimal complete sequence and solving the construction
problem for this DP will give this optimal sequence.  We can construct the code
from this optimal sequence in $O(N)$ time.

There is a subtle point here which should be mentioned.  The matrix $M$ defined by
$$
M_{i,j} = \left\{
\begin{array}{ll}
H(j) + S_{ri-j}  & \mbox{ if }  \max\{0,ri-N\} \le j < i\\
\infty  & \mbox{ otherwise}
\end{array}
\right.
$$
is Monge (the proof is similar to that of Lemma \ref{lem:c.qi}).  We can {\em not} use the SMAWK algorithm to
find its row minima and solve the problem,  though.  The reason is that,
as stated in Lemma \ref{lem:SMAWK},  the SMAWK algorithm
requires  being able to calculate any arbitrary requested entry $M_{i,j}$ in $O(1)$ time.  In
our current DP, though,  the $M_{i,j}$ are dependent upon the values $H(j)$
which are the  row-minima of other rows
in the same matrix!  Thus,  we have no way of calculating
$M_{i,j}$ in $O(1)$ time when required 
and  the SMAWK algorithm can not be applied.  This is the reason why
Larmore and Przytycka  \cite{larmore:1163}   needed to use the more sophisticated
CLWS algorithm of  \cite{Wi88} to solve the binary ($r=2$) version of this
problem.  Other algorithms for more generalized versions of the 
CLWS  have since appeared, e.g., \cite{LaSc91}, that could also be used to
solve this problem in $O(n)$ time, but they are also quite complicated.
To summarize,  by transforming $r$-ary
Huffman coding into a DP and using sophisticated tools such as
\cite{Wi88} or \cite{LaSc91} we can solve the problem in
$O(n)$ time.  This is not of practical interest, though,  since the simple, greedy, Huffman encoding algorithm is just as fast. Where the DP formulation helps is in the LLHC problem, exactly where the greedy procedure fails.  In that case we have the added practical benefit of being able to use the
simple SMAWK algorithm rather than the more complicated \cite{Wi88} or \cite{LaSc91}.

\bibliographystyle{IEEEtran}


%

\appendices
\section{Derivation of the LLHC Dynamic Program}
\label{app:LLHCDP}

In order to make this note self-contained we provide  a brief derivation of
the DP that models the LLHC.  To the best of our knowledge,  the derivation
for  the general $r$-ary case has never been written down before (although it
is known as ``folklore'').

A set of $n$ prefix-free codes in an $r$-ary alphabet can be represented by an $r$-ary tree with $n$
leaves.
The $i^{\mbox{th}}$  edge from an internal node to its children is  labeled by $\sigma_i$.
Each leaf corresponds to a code word, which is the concatenation of the characters on the
root-to-leaf path.
Then, the expected code length equals the weighted external path length of the tree.

Denote  the height of the tree by $h$.
The lowest leaves are on level $0$; the root is at level $h$.
Optimal (min weighted external path-length)
assignments of the probability $p_i$'s to the leaves  always assign smaller
probabilities to leaves at lower levels.
Since the probabilities are given in sorted order, this assignment can
be done in $O(n)$ time for a given tree.
The \emph{cost} of a tree is its weighted external path length w.r.t.~an optimal assignment.

Define the \emph{degree} of a node to be the number of its children.
A  node is \emph{complete} if it is of degree $r$, and a tree is \emph{complete} if all its
internal nodes are complete.
The following properties are easy to prove

\begin{property}
\label{pro:level.above.2}
In an optimal tree, the internal nodes at levels $\ge 2$ are complete.
\end{property}
\begin{property}
\label{pro:level.1}
There is an optimal tree that has at most one incomplete internal node, and if this node
exists, it is  at level $1$. Furthermore,
the degree of this incomplete node is $\ge 2$.
\end{property}

These properties imply that the optimal tree is almost complete and
has $\lceil \frac{n-1}{r-1} \rceil$ internal nodes.
If $n-1$ is divisible by $r-1$, the tree is complete.
Otherwise, we can add 
$$n-1 - \left\lfloor \frac{n-1}{r-1} \right\rfloor (r-1) \le r-2$$
dummy leaves to make it complete.
We assign dummy $p_i$'s with zero values to these dummy leaves.
It is easy to see that the new tree with these  dummy leaves is precisely an optimal tree for the
probabilities with the added zero-valued dummy $p_i$'s.
So, finding an optimal tree for probabilities with these dummy $p_i$'s is equivalent to the original
problem.
Therefore, w.l.o.g., we assume in the original problem, the optimal tree is a complete tree, i.e.,
we assume $n-1$ is always a multiple of $r-1$.
In this way we transform the $r$-ary Huffman coding problem to the problem of finding an optimal
complete $r$-ary tree with $n$ leaves.

A complete tree of height $h$ can be fully represented by a sequence $(i_0, i_1, \ldots, i_h)$,
where $i_k$ denotes the number of internal nodes at levels $\le k$.
Note that from this sequence  we can calculate $I_k = i_k - i_{k-1}$, the number of internal nodes \emph{on} level
$k$ and with that information we can reconstruct the tree in $O(n)$ time as follows:

\medskip
\par\noindent{}\underline{\tt Create} \\
1. For $k=1$ to $h$ \\
2. \hspace*{.1in} Create $I_k$ nodes $V_k = \{v_1,\ldots,v_{I_k}\}$ on level  $k$; \\
3. \hspace*{.1in} Create $rI_k - I_{k-1}$ leaves on level $k-1$; \\
4. \hspace*{.1in} Make $\{v_1,\ldots,v_{I_k}\}$ the parents of the $r I_k$ nodes on \\
\hspace*{.25in} level $k-1$.
\medskip

We will now see how to rewrite the cost of a tree using its representative sequence:
\begin{lemma}
\label{lem:leaf_no}
If $\cali = (i_0, i_1, \ldots, i_h)$ represents tree $T$, then $T$ has $r i_k - i_{k-1}$ leaves
on levels $< k.$
\end{lemma}
\begin{IEEEproof}
Consider the forest which is the portion of $T$ on or below
level $k$.  It is composed of  $I_k = i_{k} - i_{k-1}$ trees with roots on level $k$,

In total,  the forest contains
 $i_k$ internal nodes.

If $T'$ is a complete $r$-ary tree with $m$ internal nodes then $T'$ has $(r-1)m+1$ leaves so our forest must contain
 $(r-1) i_k + I_k = r i_k - i_{k-1}$ leaves.
\end{IEEEproof}

Recall that  $S_m = \sum_{i=1}^m p_i$ for $1 \le m \le n$.
Using the lemma above,  we have  
\begin{lemma}
\label{lem:eqn:cost.tree.2}
If the sequence $(i_0, i_1, \ldots, i_h)$ represents a tree, then the cost of the tree is
$\sum_{k=1}^h S_{r i_k - i_{k-1}}$.
\end{lemma}
\begin{IEEEproof}
Recall from Lemma \ref{lem:leaf_no} that $r i_k - i_{k-1}$ is the number of leaves at levels $< k$. So
\begin{eqnarray*}
&&\mbox{Cost} \mbox{ of the tree}\\
& = & \textstyle{\sum_{\ell=0}^{h-1} \mbox{(sum of weights of leaves at level $\ell$)} \cdot (h-\ell)} \\
& = & \textstyle{\sum_{\ell=0}^{h-1} \mbox{(sum of weights of leaves at level $\ell$)} \cdot \sum_{k=\ell+1}^h 1} \\
& = & \textstyle{\sum_{k=1}^h \sum_{\ell=0}^{k-1} \mbox{(sum of weights of leaves at level $\ell$)}} \\
& = & \textstyle{\sum_{k=1}^h \mbox{(sum of the weights of leaves at levels $< k$)}}\\
& = & \sum_{k=1}^h S_{r i_k - i_{k-1}}
\end{eqnarray*}
\end{IEEEproof}

For a complete $r$-ary tree with $n$ leaves, we have
$0 = i_0 < i_1 < \cdots < i_h = \frac{n-1}{r-1}$ and, from Lemma \ref{lem:leaf_no},  $r i_k - i_{k-1} \le n$ for all
$1 \le k \le h$. %


For technical reasons,
because we will be dealing with trees   having height {\em at most}
(but not necessarily  equal to) $h$,
we allow initial padding of the sequence by
{\bf 0}s so that a  sequence representing a tree will be of the form
$(i_0, i_1, \ldots, i_h)$ that has the following properties

\begin{definition}
\label{def:valid.list.2}
A sequence $(i_0, i_1, \ldots, i_h)$ 
is a valid $(n,r)$-sequence, if
\begin{itemize}
\item $\exists t$ such that $i_0= i_1 = \cdots=i_i=0.$
\item  $0 < i_t < \cdots < i_h \le  \frac{n-1}{r-1}$
\item  $r i_k - i_{k-1} \le n$ for all $1 \le k \le h$.
\end{itemize}
A sequence  is {\em complete} if it is valid and $i_h = \frac{n-1}{r-1}$.
\end{definition}
It is straightforward to see that padding the sequence representing a tree with initial {\bf 0}s,
does not change the tree built by the {\tt Create} procedure or the validity of Lemmas
\ref{lem:leaf_no} and \ref{lem:eqn:cost.tree.2}.

We can now  extend our cost function to {\em all}  valid $(n,r)$-sequences
 sequences, not just the ones representing trees.

\begin{definition}
\label{def:cost.2}
For valid  $(n,r)$-sequence $\cali = (i_0, i_1, \ldots, i_h)$ define
$$cost(\mathcal{I}) = \sum_{k=1}^h S_{r i_k - i_{k-1}}.$$

$\cali$ is {\em optimal} if
$cost(\cali) = \min_{\cali'} cost(\cali')$ where the minimum is taken over all valid length $h$
 $(n,r)$-sequences
$\cali' = (i'_0, i'_1, \ldots, i'_h)$ with $i'_h = i_h,$ i.e., all sequences of the same length that end with
the same value.
\end{definition}
{\small \em
Note: padding a sequence with initial {\bf 0}s doesn't change
its completeness or cost. Furthermore,  if $\cali$ is created by padding
the sequence corresponding to tree $T$ with initial {\bf 0}s,  then procedure
{\tt Create} will still recreate $T$ from $\cali.$
}

\medskip

It follows from the definitions that
for fixed $(n,r)$ we can calculate $H(d,j)$, the cost of
an optimal  $(n,r)$-sequence $(0,i_1,i_2,\ldots,i_d)$ with $i_d=j$ using the DP
(\ref{eq:DP_Def}) with
\begin{equation}
\label{eq:defc_2}
c_{i,j}^{(d)} =
\left\{
\begin{array}{ll}
0         & \textrm{if } $i=j=0$\\
S_{ri-j} & \textrm{if } \max\{0,ri-n\} \le j < i \\
\infty & \textrm{otherwise.}
\end{array}
\right.
\end{equation}

The subtle issue is that not all complete sequences correspond to trees, e.g,
$(0,3,4,5)$ is a complete $(6,2)$ sequence that does not represent any binary tree.
Thus, a-priori, finding an optimal complete sequence might not help us find an optimal tree.
We are saved by the next lemma.
\begin{lemma}
\label{lem:sequence.2.tree.2}
An optimal complete  $(n,r)$-sequence always represents a tree.
\end{lemma}
Thus, we can find an optimal tree by first solving the construction problem
for  DP (\ref{eq:DP_Def}) with conditions (\ref{eq:defc_2}) to get
an optimal complete $(n,r)$-sequence $\cali$  and then
building the tree that corresponds to $\cali$.

\begin{figure*}[!htbp]
\centering
\includegraphics{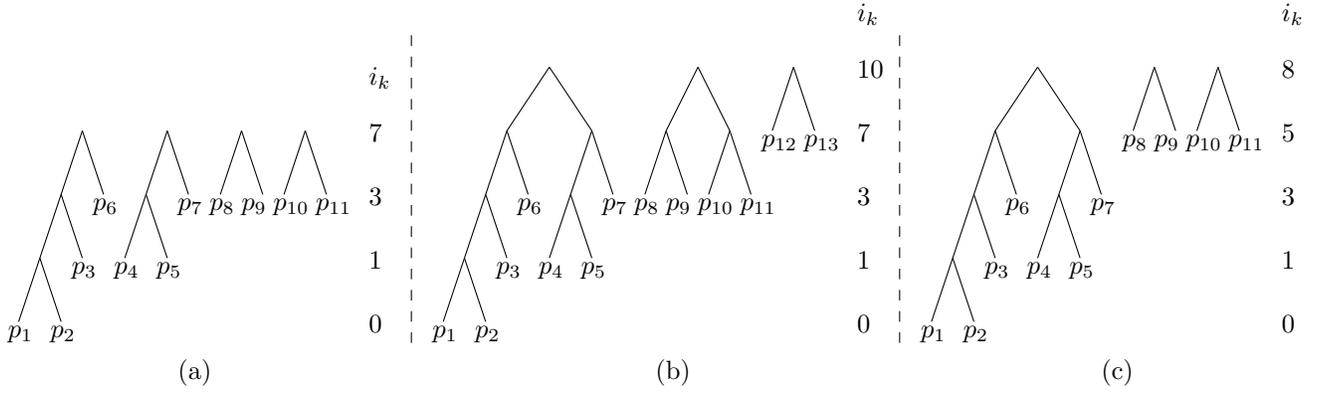}
\caption{Illustration of the two cases in the proof of Lemma \ref{lem:sequence.2.forest.2}.
Here, $r = 2$ and $h = 4$.
(a) is the forest $F'$ corresponding to the old sequence $\cali' = (0,1,3,7)$.
(b) illustrates  case $1$: if $i_h=10$ then $I_h = 3$ and $2 I_h = 6 \ge 4 = I_{h-1}$ so
we can create a forest corresponding to the new sequence  $(0,1,3,7,10)$.
(c) illustrates  case $2$: if $i_h=8$ then $I_h = 1$ and  $2 I_h =2 < 4 = I_{h-1}$.  In this case
the sequence $\bar \cali = (0,1,3,5,8)$ (corresponding to the forest pictured) has cost
$S_2 + S_5 + S_7 + S_{11}$.  This is cheaper than the cost $S_2 + S_5 + S_{11} + S_9$ of 
 the sequence $\cali = (0,1,3,7,8)$. As noted in the proof,
$\bar \cali$ is constructed
by lifting   two subtrees in the forest in (a) and then writing down the corresponding sequence.}
\label{fig:forest}
\end{figure*}

Before proving Lemma \ref{lem:sequence.2.tree.2} we will need to extend our definitions
from trees to forests. See Figure \ref{fig:forest}(a).
\begin{definition}
A legal $(n,r)$-forest, or {\em forest},  is a collection of complete $r$-ary trees that together contain
at most $n$ leaves, all of whose roots are at the same height.
\end{definition}
Given $p_1 \le p_2 \le \ldots,p_n$ we can assign the $p_i$ to the leaves of forest $F$
from bottom to top of tree and define the cost of $F$ (with respect to the $p_i$) to
be the sum  of the  costs of its component trees.
Note that a tree with $n$ leaves is a forest and its cost as a forest will be the
same as its cost as a tree.

Now, for forest  $F$ let
$i_k$ be the number of internal nodes it has at level $\le k$.  Then,  we can talk
about the sequence $\cali = (i_0,i_1,\ldots,i_h)$ associated with the forest.  Reviewing
the proofs of Lemmas  \ref{lem:leaf_no}  and \ref {lem:eqn:cost.tree.2} we see that they
were actually statements about forests and not trees so
$F$ has $r i_k - i_{k-1}$  leaves on levels $<k$ and
$cost(F) = cost(\cali).$

We will  prove
\begin{lemma}
\label{lem:sequence.2.forest.2}
An optimal $(n,r)$-sequence $\cali=(i_0,i_1,\ldots,i_h)$ always represents a forest.
\end{lemma}

Note that this will immediately imply  Lemma \ref{lem:sequence.2.tree.2} because if $\cali$ is complete then
$i_h = \frac {n-1} {r-1}$
and,  by validity, $r i_h - i_{h-1} \le n$, implying $i_{h-1} = i_h -1$.
Thus the forest corresponding to $\cali$ is composed of exactly $i_h - i_{h-1} =1$ trees at level
$h$ and
is therefore a tree itself.

\begin{IEEEproof} (of Lemma \ref{lem:sequence.2.forest.2})\\
Without loss of generality assume that $i_0 =0 < i_1$.
Our proof will be by induction on $h$.

First note that if $h=1$, then $\cali = (0,i_1)$ for some $i_1 >0$
and this represents the forest composed of $i_1$ complete trees each
of height $1$ so the lemma is trivially correct.

Now let $h >1$.
Set $I_h = i_h - i_{h-1}$ and $I_{h-1}= I_{h-2}-I_{h-1}.$

Define $\cali'=(i_0,i_1,\ldots,i_{h-1}).$  Since
$\cali'$ is optimal,
by induction,  $\cali'$ represents a forest $F'$
with $I_{h-1}$ roots at level $h-1$ and a total of  $L_{h-1} = r i_{h-1} - i_{h-2}$ leaves.
There are now two cases: see Figure \ref{fig:forest}.

\par\noindent\underline{Case 1:  $r I_h \ge I_{h-1}$}:\\
Then $\cali$ represents a forest with $I_h$ roots whose $r I_h$ children are
exactly the $I_{h-1}$ roots from $F'$ and another $r I_h - I_{h-1}\ge 0$ leaves.  So the Lemma is correct.

\par\noindent\underline{Case 2:  $r I_h < I_{h-1}$}:\\
We will show that this contradicts the optimality of $\cali$ and is therefore impossible.
Thus Case 1 will be the only possible case and the Lemma correct.

Assume now that  $r I_h <  I_{h-1}$   and set  $s = I{h-1} - r I_{h} > 0$.
This can be rewritten as
$r (i_h - i_{h-1})  + s = r (i_{h-1} - i_{h-2})$ so
$$ r i_h - i_{h-1} =  r i_{h-1} - i_{h-2} -s = L_{h-1} -s .$$

Now consider $F$ as being labeled with the $L_{h-1}$ smallest $p_i$ and
construct a new forest $\bar F$ as follows.
Choose $s$ trees from $\bar F$ containing the $s$ largest weights in the forest, i.e.,
$p_j,$
$j=L_{h-1}, L_{h-1}-1,\ldots, L_{h-1} - (s-1).$
Move those $s$ forests up one level so their roots are now at height $h$ and not $h-1.$
Now add $I_h$ new nodes to level $h$.  Make them the parents of the remaining $r I_h$ nodes on level $h-1.$
This forest is a legal forest.  Call its representative sequence $\bar \cali = (\bar i_0,\bar i_1,\ldots,\bar i_h).$

We now observe \\
(a)  $\bar i_{h-1} = i_{h-1} - s$ so
$$\bar i_h = \bar i_{h-1} + s + I_{h} = i_{h-1}  + I_{h} + s =  i_h.$$
(b) Thus $r \bar i_h - \bar i_{h-1} = r  i_h - (i_{h-1} -s) = L_{h-1}$  and
$$S_{r \bar i_h - \bar i_{h-1}} =
S_{L_{h-1}}
= S_{r i_h - i_{h-1}} +  \sum_{j= L_{h-1} -s  +1}^{L_{h-1}}  p_j
$$
(c) Let $\bar F'$   be levels $0$-$(h-1)$ of $\bar F$.  Since every complete tree contains
at least $r$ nodes, the $s$ trees raised contain at least
the $s$ nodes $p_j$ where $ L_{h-1} -s  < j \le L_{h-1}$ and one other node.  Since every such node was raised one level,
\begin{eqnarray*}
\sum_{m=1}^{h-1} S_{r\bar i_m - \bar i_{m-1}}
&=& cost(\bar F')\\
&<& cost (F') - \sum_{j= L_{h-1} - s +1}^{L_{h-1}}  p_j\\
&=& \left(\sum_{m=1}^{h-1} S_{r i_m - i_{m-1}} \right) -  \sum_{j= L_{h-1} -s  +1}^{L_{h-1}}  p_j
\end{eqnarray*}
Combining (b) and (c) shows that $cost(\bar \cali) < cost(\cali)$.  This is a contradiction since
both $\cali$ and $\bar \cali$ are valid sequences of length $h$ that end with the same value $i_h$
and $\cali$ is optimal.  Thus the case $r I_h <  I_{h-1}$  can not happen and we are finished.

\end{IEEEproof}

\ifCLASSOPTIONcaptionsoff
  \newpage
\fi

\end{document}